\documentclass[aps,pra,superscriptaddress,twocolumn,10pt]{revtex4-1}

\usepackage{graphicx}
\usepackage{bm}
\usepackage[usenames,dvipsnames,svgnames,table]{xcolor}
\usepackage{epsfig}
\usepackage{amsmath}
\DeclareMathOperator{\Tr}{Tr}

\usepackage{amssymb}
\usepackage{array}
\usepackage{nccmath}
\usepackage{dsfont}
\usepackage{array}
\usepackage{cancel}
\usepackage{hyperref}
\hypersetup{colorlinks=true,breaklinks,linkcolor=blue,urlcolor=blue,citecolor=blue}

\usepackage{ulem}

\allowdisplaybreaks
\begin{abstract}
   In the present paper, a global Linbladian ansatz is constructed which leads to thermalization at temperature $T$ to the Gibbs state of the investigated system.
   This ansatz connects every two eigenstates of the Hamiltonian and leads to a simple master equation known in the literature as the relaxation time approximation (RTA).  
   The main message of this paper is that RTA, being a Libladian approach itself,  can be used as Linbladian securing thermalization when modeling physical processes, and can be consequently combined with other types of Linbladians which would drive the system of the equilibrium state.
   \newline
   I demonstrate it with two applications. The first application is the slow cooling (or heating) of quantum systems by varying the environment temperature to a critical point. With this RTA-Libblad ansatz, one can directly relate to the equilibrium behavior of the system, and if an order parameter has the exponent $\Psi$, the remaining value at the phase transition will decrease with $1/\tau^{\Psi}$, where $\tau$ is the overall time of the slow process.
   \newline
   In the second application, I investigate the change in the expectation value of a conserved quantity (an operator commuting with the Hamiltonian) due to an extra Linbladian term which would drive the system out from equilibrium, while the thermalizing RTA-Linbladian term is also present. I give a closed perturbative expression in the first order for the expectation value in the new steady state using only expectation values calculated in the original thermal equilibrium.
\end{abstract}

\begin{document}
\title{Linbladian way for the relaxation time approximation, application to Kibble - Zurek processes due to environment temperature quench, and to Linbladian perturbation theory}
\author{Gerg\H o Ro\'osz} 
\affiliation{%
    HUN-REN Wigner Research Centre for Physics, H-1525 Budapest, P.O.Box 49, Hungary}%
\maketitle

Recently there has been a growing interest in open many-body quantum systems \cite{Zala_GGEs_2018,Arrigoni2013,MarkoZ2015,Temme2013LowerBT,Strathearn2018, THOMPSON2023169385, Prosen_2008, Dzhioev_2012,THOMPSON2023169385} due to their importance in quantum information processing experiments where the isolation from external noise and decoherence is essential \cite{MAX_SCHLOSSHAUER2019, Rezvani2021, WANG2024, Weimer2021,Hermann2005, MI_science, Albert2016}, and it is important to understand how the experiment setup interact with the environment.


One way to describe the system-environment interaction is Lindblad formalism, which physically corresponds to a Markovian approximation in the weak coupling limit \cite{Marco2024}.
 
When describing a physical system in a Markovian way in the Lindblad formalism, finding the proper Lindblad operators is nontrivial \cite{Wang2024-master-construct, Babu2024, Dann2021quantumthermo}.

One formal solution is, to couple locally \cite{two-site-hubbard} at the boundaries to heat reservoirs. 
While local couplings are well suited to describe non-equilibrium steady states \cite{Arrigoni2013,consistent_local_De_Chiara_2018,gonzalez-global-vs-local,PhysRevE.107.014108-consistent-local-master-equation,Shinkov2020}, in other processes, such as during the cooling of a system \cite{King-univ-cooling-PhysRevLett.130.050401,Bacsi2023-ising,Juhasz-PhysRevB.108.224203} the global coupling to an environment is more physical. 
Many works describing open quantum systems globally coupled to a finite temperature reservoir introduce the coupling to the environment using explicitly the eigenvalue decomposition of the Hamiltonian \cite{Bacsi2023-ising,bacsi2023lindbladian-Luttiger,King-univ-cooling-PhysRevLett.130.050401, Juhasz-PhysRevB.108.224203}.
While this global coupling is compatible with thermodynamics\cite{cattaneo2019local}, for many interesting models the eigenvalue decomposition is unknown. Even when reliable numerical solutions exist, for example quantum Monte-Carlo, or Feynman diagrams, one usually avoids to describe the spectrum of a many-body system and investigates directly expectation values instead.

There is a recent paper  where the authors construct a global Lindbladian description, which can be used without explicit knowledge of the spectrum \cite{Universal_Lindblad_equationPhysRevB.102.115109}. 
In \cite{Universal_Lindblad_equationPhysRevB.102.115109} strong or intermediate interactions are allowed, which is an important step forward; however, it leads to steady states different from the thermal equilibrium state, which is physically relevant if the system-reservoir interaction is strong, but when modeling an almost isolated system, or slow cooling of quantum systems, one wants an ansatz whose steady state is the thermal equilibrium (Gibbs) state.
So a global Limbladian ansatz, which can be used without detailed information on the spectrum of the Hamiltonian, and relaxing to a Gibbs state would be useful.

In this work, I show that the well-known relaxation time approximation (RTA) can be derived from a Linbladian ansatz, and in this way, it can be viewed as a global Linbladian whose steady state is the Gibbs state, and in many cases, one does not need the spectrum to handle it.

The paper is organized as follows. In the first section, I give the Liblad ansatz for the RTA. In Section II. I apply it to environment temperature quenches. In Section III. a first-order perturbation formula is derived for integrals of motions, in Section IV. it is applied to a fermion chain.

\section{Lindblad ansatz for the RTA}
I start with the eigenvalue decomposition, connect every two eigenstates with Lindblad generators, and then choose the coefficients to ensure the thermal equilibrium  density matrix 
\begin{equation}
	\rho_E(\beta) = \frac{e^{-\beta H}}{Z}
\end{equation}
is a stationary solution.

I search the Lindblad operators in the following form
\begin{equation}
	L_{i,j} = C_{i,j} | i \rangle \langle  j |
\label{eq:gen}
\end{equation}
where the $| i \rangle$ vectors  denotes the eigenvectors of the Hamiltonian and the $C_{i,j}$ constants are unknown. We allow diagonal operators   
\begin{equation}
	L_{i,i} = C_{i,i} | i \rangle \langle  i |
\end{equation}
due to technical reasons. These operators correspond to added white noise \cite{Gergo_2016, Armin2015}.
The Lindblad equation is
\begin{equation}
	\frac{\partial \rho}{ \partial t} = -i [H,\rho] + \sum_{i,j} L_{i,j} \rho L^\dagger_{i,j} - \frac{1}{2} \left\{ L^\dagger_{i,j}  L_{i,j}, \rho \right\}
   \label{eq:lindblad_gen}
\end{equation}
Substituting E.q. \ref{eq:gen} and collecting the terms, one got

\begin{align}
	&0=\sum_{i,j} (C^2_{ij} e^{-\beta E_j} - C^2_{ji} e^{-\beta E_i}) | i \rangle \langle i |  \\
	&C^2_{ij} e^{-\beta E_j} = C^2_{ji} e^{-\beta E_i}
\end{align}
I choose the Lindblad operators to be
\begin{equation}
	L_{i,j} = \sqrt{\gamma_0} \frac{e^{-\beta E_i/2}}{Z} | i \rangle \langle j|
	\label{eq:generator}
\end{equation}
where the $\gamma_0$ constant describe the strength of the coupling with the reservoir. 
In many relevant physical models, where the Lindblad generators connect different particle subspaces of the Fock space, this means that the system is connected to heat and also particle bath.
It is clear that the dynamics have the desired stationary solutions, 
and we will see that it leads to the RTA.

To see these, I substitute the generators of E.q. (\ref{eq:generator}) to the Lindblad equation (\ref{eq:lindblad_gen})  and got
\begin{align}
	\label{eq:sum_up}
	& \frac{\partial \rho}{ \partial t} = -i [H,\rho]  \\
	& +\gamma_0  \left[ \underbrace{\left( \sum_i \frac{e^{-\beta E_i}}{Z} |i \rangle \langle i | \right)}_{\rho_E(t)}  \underbrace{\left( \sum_j \langle j |\rho| j \rangle \right)}_1 \right.\nonumber \\
	& -\frac{1}{2}\underbrace{\left( \sum_i \frac{e^{-\beta E_i}}{Z}  \right)}_{1} \underbrace{\left( \sum_j \langle j |\rho| j \rangle \right)}_1 \rho \nonumber \\ 
	& \left.  - \frac{1}{2}\rho \underbrace{\left( \sum_i \frac{e^{-\beta E_i}}{Z}  \right)}_{1} \underbrace{\left( \sum_j \langle j |\rho| j \rangle \right)}_{1}  \right]  \nonumber
\end{align}
The  closed form of the equation is
\begin{equation}
\frac{\partial \rho}{\partial t} = i [H,\rho] + \gamma_0 \left( \frac{e^{-\beta H}}{Z_0} -  \rho \right) 
\label{eq:compact_form}
\end{equation}
 Here E.q. \ref{eq:compact_form} is the well-known RTA equation.
 The RTA has a constant term in it, while the usual form of the Lindblad equation does not. We have seen, that the constant emerges when we summed up a lot of terms in E.q. (\ref{eq:sum_up}).
 I would like to emphasize that the relaxation time approximation is often used to describe quantum systems \cite{REINHARD2015183,Strickland2019,DINH2022}, however the connection to the Lindbladian formalism is usually not mentioned.
\section{Environment temperature quench}
I consider here a slow, almost adiabatic change of the environment temperature.  
The first process is the slow cooling to  zero temperature
\begin{eqnarray}
	T(t) = T_0 \left(1-\frac{t}{\tau}\right)
\end{eqnarray}
which is $T_0$ initially and decreases to zero temperature in time $\tau$.
The other process is slow heating, where the system is slowly heated from low temperatures to the critical point as

\begin{equation}
    T(t) =T_{crit}\frac{t}{\tau}
\end{equation}

The E.q. The master equation ( \ref{eq:compact_form}) is modified so that the equilibrium density matrix is now replaced by the equilibrium density matrix at time $t$.
\begin{equation}
	\rho_E(t)=\frac{\exp(-\beta(t) H )}{Z(t)}
\end{equation}

\begin{equation}
	\frac{\partial \rho}{\partial t} = i [H,\rho] + \gamma_0 \left( \rho_E(t) -  \rho \right) 
	\label{eq:compact_form_kibble_zurek}
\end{equation}
One finds the following closed (exact) solution
\begin{equation}
	\rho(t) = \frac{e^{-\beta(0) H }}{Z(0)} e^{-\gamma_0 t} + \gamma_0 \int_0^t dt' e^{\gamma_0 (t'-t)} \frac{e^{-\beta(t')H}}{Z(t')}
	\label{eq:solution}
\end{equation}
The expectation value of a given operator $O$ is 
\begin{equation}
	\langle O \rangle (t) = e^{-\gamma_0 t} \langle O \rangle_{T_0} + \gamma_0 \int_0^t dt' e^{\gamma_0 (t'-t)} \langle O \rangle_{T(t)} 
\label{eq:operator_expectation_value}
\end{equation}
Here I would like to remark, that the previous studies in quadratic fermion models have written up E.q. (\ref{eq:operator_expectation_value}) in the special case of the fermionic occupation number \cite{Bacsi2023-ising,Juhasz-PhysRevB.108.224203,King-univ-cooling-PhysRevLett.130.050401}, and a similar solution appears in the literature of the semi-classical RTA description \cite{Strickland2019}.



Let's assume that the system has a finite temperature phase transition at $T_{crit}$, the order parameter $m$ is nonzero for $T<T_{crit}$ and decreases near the phase transition as
\begin{equation}
    \langle m \rangle \sim \left( \frac{T_{crit}-T}{T_{crit}} \right)^\Psi
\end{equation}
Using E.q. \ref{eq:operator_expectation_value} one got for the 

\begin{align}
\langle m \rangle (t) &= \int_0^t e^{-\gamma_0 (t-t')} \left( \frac{T_{crit}-t'/\tau}{T_{crit}} \right)^\Psi\\
\langle m \rangle (\tau)&\sim\frac{1}{\tau^{\Psi}}
 \end{align}
 Here we neglected the exponentially small first term.
In the other process, heating up to a finite temperature critical point one finds an identical scaling. In Table \ref{tab:scalings} I list the scaling of the relevant quantities in some well-known physical systems.
 
 \begin{table}
     \centering
     \begin{tabular}{|c|c|}
        \hline heating up Bose &  \\
         condensation in a box &$N_0 \sim 1/\tau^3$ \\ \hline
          heating up Bose& \\
          condensation in 3D harmonic trap & $N_0 \sim 1/\tau^2$ \\ \hline
         cooling non-interacting 3D & \\
          Fermions to $T=0$ & $\delta E \sim 1/\tau^2$ \\ \hline
         cooling homogeneous transverse   & \\
        Ising chain to $T=0$ &  $N_f\sim 1/\tau^2$ \\ \hline
            cooling disordered transverse   & \\
        Ising chain to $T=0$ &  $N_f\sim 1/\ln^2(\tau)$ \\ \hline
     \end{tabular}
     \caption{Relaxation time approximation results for the residual condensation particle numbers when heating up Bose condensations from $T=0$ to the critical temperature and for the residual energy when slowly cooling free fermions to $T=0$ and the number of remaining fermionic excitations in clean or disordered critical quantum transverse Ising chains cooled to to $T=0$.}
     \label{tab:scalings}
 \end{table}

\section{Perturbation for the expectation values of integrals of motion}
Now I assume that the system is initially in equilibrium at temperature $T$, and it is connected to an environment of temperature $T$ acting in the RTA way.
Now, one turns an extra Linbladian term on, which would drive the system of equilibrium. The new steady state will be the result of the two competing processes.
\begin{equation}
\frac{\partial \rho}{\partial t} = i [H,\rho] + \gamma_0 \left( \frac{e^{-\beta H}}{Z_0} -  \rho \right) + \epsilon {\cal L}_1 [ \rho ]
\label{eq:pert_time_ev}
\end{equation}
Where 
\begin{equation}
    {\cal L}_1 [ \rho ] = \sum_{j=1}^n L_{(1),i} \rho L^\dagger_{(1),i} -\frac{1}{2} L^\dagger_{(1),i} L_{(1),i} \rho -\frac{1}{2} \rho L^\dagger_{(1),i} L_{(1),i} \;.
\end{equation}
The subscript 1 here stands to make a difference between the notation used to construct the RTA-Lindblad ansatz, and the extra Lindbladian terms introduced here. 
Here I assume that the disturbance is much weaker than the thermalization process $\epsilon \ll \gamma_0$.
The equilibrium state is characterized by
\begin{equation}
0 = i [H,\rho] + \gamma_0 \left( \frac{e^{-\beta H}}{Z_0} -  \rho \right) + \epsilon {\cal L}_1 [ \rho ]
\label{eq:pert_time_ev}
\end{equation}
One write for the difference between the new steady state density matrix and the original thermal equilibrium
\begin{equation}
    \rho-\rho_0 = \frac{\epsilon}{\gamma} {\cal L}_1[\rho] + \frac{i}{\gamma} [H,\rho]
    \label{eq:formal_rho}
\end{equation}
Now I consider an integral of motion $O$, 
\begin{equation}
    [O,H] =0
\end{equation}
 To get the change in the expectation value of $O$ due to the new Lindbladian term one multiplies \ref{eq:formal_rho} with $O$ and takes the trace of bought sides.
 \begin{equation}
     \delta \langle O \rangle =  \frac{\epsilon}{\gamma} \Tr( O {\cal L}_1[\rho]) + \frac{i}{\gamma}\Tr (O[H,\rho])
     \label{eq:small_cange_in_O}
 \end{equation}
 In E.q. \ref{eq:small_cange_in_O} the second term is zero $\Tr (O[H,\rho])= \Tr (O[H,\rho]) = \Tr ( \rho[H,O])=0$. In the first term,   the new equilibrium state $\rho$  is unknown.  However, the disturbance is small, and the difference between the new equilibrium state and the thermal equilibrium is also small $\delta \rho \sim \epsilon/\gamma$, so in leading order one got the right value if substitute $\rho_0$ for $\rho$ in the first term of E.q. (\ref{eq:small_cange_in_O}). In this way on got the following perturbative formula for the expectation value of $O$:

 \begin{equation}
     \delta \langle O \rangle =  \frac{\epsilon}{\gamma} \Tr( O {\cal L}_1[\rho_0]) 
     \label{eq:perturbation_result_shorthand}
 \end{equation}

 \begin{align}
     &\delta \langle O \rangle = \frac{\epsilon}{\gamma}\sum_{j=1}^n  \left[ \langle L^\dagger_{(1),i} O L_{(1),i} \rangle_0 \right. \nonumber \\
     &-\frac{1}{2} \langle O L^\dagger_{(1),i} L_{(1),i} \rangle_0
     -\frac{1}{2} \langle L^\dagger_{(1),i} L_{(1),i} O \rangle_0 \left. \right]
 \label{eq:perturbation_result_detail}
 \end{align}
 Where $\langle \dots \rangle_0 = \Tr[\rho_0 \dots ] $ denotes the expectation value in the thermal average.
 This perturbation theory is motivated by \cite{Zala_2018_pert}, however, I investigate here a different setting, in \cite{Zala_2018_pert} the authors do not assume a general thermalization term (RTA in my work) which bonds the system near to the equilibrium, their approach is more flexible, they concentrate on conserved quantities but they perturbation theory can be formulated for any operator. The formula E.q. (\ref{eq:perturbation_result_detail}) I derived here holds only for conserved quantities, however, the benefit is that it is explicit, and can be computed if one is able to compute (exactly or approximately) equilibrium expectation values, and do not need the full solution of the non-perturbed problem as other formulations of Lindbladian perturbation theory do \cite{Zala_2018_pert,Li2014}.
 
 In the next section, an example is shown of how to apply E.q. (\ref{eq:perturbation_result_detail}).
 \section{Free fermion chain coupled locally to a zero temperature reservoir and globally to finite temperature reservoir}
 In this section, I consider a spinless free fermion chain length of $L$ with fee boundary conditions
 \begin{equation}
     H=-t \sum_{l=1}^{L-1} c^{\dagger}_{l+1} c_l +  c^{\dagger}_l c_{c+1}
 \end{equation}
 where $\{ c_i, c^\dagger_j \}  = \delta_{i,j}$ are fermionic creation-annihilation operators.
 The system is globally coupled to a finite temperature reservoir with the RTA-Lindblad ansatz, and site $l$ is locally coupled to a zero temperature reservoir according to the following Linbladian operators \cite{hubbard-relax}
 \begin{align}
     L_{(1),AN} &= c_l n_l  \\
     L_{(1),CR} &= c^\dagger_l (1-n_l)
 \end{align}
 The prefactors of the operators are $\epsilon_{AN}$, $\epsilon_{CR}$ respectively.
 This operator (even without the RTA-Lindblad term) breaks the integrability of the model with the superoperator method \cite{Prosen_2008}, the situation corresponds to the "Markovian integrability breaking" discussed in \cite{Lange2017}.
 The closed system is diagonalized 
 \begin{align}
     c_k &= \sqrt{\frac{2}{L}} \sum_{n=1}^L \sin\left( \frac{\pi k}{L+1} n \right) c_n\\
     H &=-2t \sum_{k=1}^L \cos \left( \frac{\pi k}{L+1} \right) c^\dagger_k c_k 
 \end{align}
 where $k = 1 \dots \L$. 
 The conserved quantities we are investigating are the fermionic number operators 
 \begin{equation}
     n_k = c^\dagger_k c_k
 \end{equation}
 With simple calculations, one got the change of the occupation numbers
 \begin{align}
     \delta \langle n_k \rangle &=   \epsilon_{CR} \frac{2}{L} \sin^2 \left( 
\frac{\pi k}{L+1} l\right) (1-\langle n_k \rangle) \\
&- \epsilon_{AN} \frac{2}{L} \sin^2 \left( \frac{\pi k}{L+1} l \right)\langle n_k \rangle 
 \end{align}
\section{Concluding remarks}

The mean field solution of physical systems can be understood as investigating the system on a full graph rather than a lattice. This usually leads to simplifications, but one still gets usable information about the behavior of the model over the upper critical dimension.

In this work the spatial lattice of the models is not altered, however, the Lindbladian operators are chosen so that they connect directly all states with all other states, and the connectivity graph of the states is a complete graph.  As one can see, it leads to a relevant simplification and equal relaxation time for all measurable and gives a Lindbladian ansatz to the relaxation time approximation.

 In this paper I showed that one can relate the solution of the slow cooling or slow heating problems directly to the equilibrium behavior, one got back the free-fermion results for the Ising and Kitaev chains, and one got theoretical predictions for all models where the critical exponents are known.

As a second application, I derived a simple first-order perturbation expansion, for the case where other Linbladian terms are present and the steady state is characterized by the competition of the two different processes and illustrated it with the example of one-dimensional free fermions coupled globally (with RTA) to a finite-temperature reservoir and locally to a zero-temperature reservoir.
However, in this example, the integrability is broken by the zero-temperature reservoir, and the perturbative formula gives predictions for the change of the expectation values of the conserved quantities.

While in this case, one had an exact solution for the expectation values, I would like to emphasize, that one can use the perturbative formula with approximate or numerical values.

There are a number of models, where there are interesting integrals of motions, and an analytical solution is not known, however, there are reliable approximate methods.

For example, the compass model \cite{Janke2008} , the Toric code \cite{Vidal-toric-2009}, and certain plaquette models defined with Majorana fermions \cite{Carsten2020,Carsten2022} do have a high number of conserved quantities, but the solution is not trivial or not known.
A further application could be to combine the RTA-Lindblad ansatz with the Lindblad Floquet theory presented in \cite{chen2024}. 
 A further application of E.q. \ref{eq:perturbation_result_detail} may be to investigate the effect of continuous measurement \cite{Gammelmark2013,Gross_2018,Gammelmark_PhysRevLett} on the average time evolution of monitored systems.  The dynamics averaged over the measured values are characterized by a Lindblad equation, where the Lindblad generator is the measured Hermitian operator itself. If one select the Hamiltonian as a conserved quantity and set $\gamma_0$ to the typical relaxation time in the system then one can obtain the expectation value of the energy gain in the stationary state. Since a Hermitian Lindbladian corresponds to a white noise that typically heats up a system (when there is no other Lindbladian term in the model), the long time energy gain may be used as a measure of the sensitivity of the system to continuous measurement.

\begin{acknowledgements}
The author  has been supported by  the  QuantERA II project HQCC-101017733., National Research,
Development and Innovation Office NKFIH under Grant
No. K128989,  No. K146736,  and by the Hungarian Lóránt Eötvös mobility scholarship.
The author is grateful to Karsten Held and Enrico Arrigoni for their insightful comments and suggestions. 
\end{acknowledgements}

\bibliography{main}

\end{document}